\documentclass[12pt]{article}                            
\usepackage{graphicx}
\newcommand{\comma}{\;\;\; ,}
\newcommand{\period}{\;\;\; .}
\newcommand{\eq}{\; = \;}
\newcommand{\sep}{\;\; , \;\;\;}
\newcommand{\be}{\begin{equation}}

\newcommand{\bd}{\begin{displaymath}}
\newcommand{\sn}{{\rm \, sn \, }}
\newcommand{\cn}{{\rm \, cn \, }}
\newcommand{\dn}{{\rm \, dn \, }}

\newcommand{\ee}{\end{equation}}
\newcommand{\ed}{\end{displaymath}}
\newcommand{\ba}{\begin{eqnarray}}
\newcommand{\ea}{\end{eqnarray}}

\newcommand{\half}{\textstyle \frac{1}{2}}

\newcommand{\emm}{m}

\def\picture #1 by #2 (#3){
  \vbox to #2{
    \hrule width #1 height 0pt depth 0pt
    \vfill
    \special{picture #3}}}
\def\scaledpicture #1 by #2 (#3 scaled #4){{
  \dimen0=#1 \dimen1=#2
  \divide\dimen0 by 1000 \multiply\dimen0 by #4
  \divide\dimen1 by 1000 \multiply\dimen1 by #4
  \picture \dimen0 by \dimen1 (#3 scaled #4)}}

\title{Equivalence of the two results for the free energy
of the chiral Potts model}

\author{ R.J. Baxter\\
{\protect \small Theoretical Physics, I.A.S. and School of Mathematical
Sciences}\\
{\protect \small  The Australian National University,
 Canberra, A.C.T. 0200, Australia  }
\thanks{This work was performed while the author was a Visiting Miller Professor at the
University of California at Berkeley.}
}

\date{\today}

\begin{document}

\maketitle

\abstract{The free energy of the chiral Potts model has been obtained in two
ways. The first used only the star-triangle relation, symmetries and invariances,
and led to a system of equations that implicitly define the free energy, and from which 
the critical behaviour can be obtained. The second used the functional relations derived by
Bazhanov and Stroganov, solving them to obtain the free energy explicitly as a double
integral. Here we obtain, for the first time, a direct verification that the two
results are identical at all temperatures.
}

\section{Introduction}

The chiral Potts model is the first model found that satisfies the
star-triangle relation but does not have the ``difference property''.
This means that the model and its properties cannot be simply expressed
in terms of elliptic functions.
Its free energy was first investigated by
showing that it satisfied certain partial differential equations
\cite{freeenergy88}. Later an explicit formula was obtained as a double
integral \cite{RJB90, Seoul}, using the transfer matrix functional relations derived by
Bazhanov and Stroganov \cite{BazStr90} and discussed by the author \cite{RJBFR}. These two
results were shown to be consistent in the scaling region near criticality
\cite{RJB96}, but their equivalence has not been verified in general. Here we do this and
discuss the properties of the functions $x_p, y_p$ that play a key role in \cite{freeenergy88}.

The model is defined in Ref. \cite{BPAY88}. Here we shall consider only the model on the
square lattice, but because of the star-triangle relation corresponding models can also be
defined on the triangular and honeycomb lattices. It is shown in \cite{freeenergy88} how
their free energies can easily  be obtained from that of the square lattice.

 Here we use the
notation of
\cite{freeenergy88}. Let $k, k'$ be two real numbers, between 0 and 1, satisfying
\be k^2 + k'^2 = 1 \period \ee
Define variables
$\theta_p$, $\phi_p$, $u_p$, $v_p$,
related to the $a_p, b_p, c_p, d_p$ of \cite{BPAY88} and to one another
by
\bd
e^{i \theta_p} =  e^{-\pi i /N} \, b_p /c_p \sep
e^{i \phi_p} =  a_p /d_p \comma \ed
\be \label{thetaphi}
u_p  = N (\theta_p + \phi_p )/2 \sep v_p  = N (\theta_p - \phi_p )/2
\ee
Then from eqn. (9) of \cite{BPAY88}, $u_p$ and $v_p$ are related by
\be \label{vu} \sin v_p \eq k \, \sin u_p \period \ee

Thus if the constant $k$ is given,
then any one of the parameters $\theta_p$, $\phi_p$, $u_p$, $v_p$ specifies the
other three, to within multiple, but discrete, values. Here we shall regard 
$u_p$ as the primary (real) variable, and usually confine it to the interval

\be \label{u-interval} 0 < u_p < \pi \period \ee

Then there is a unique choice for $v_p$ such that it is real, satisfying

\be -\pi /2 < v_p < \pi /2 \period \ee
The  parameters $\theta_p$, $\phi_p$ are then uniquely defined by (\ref{thetaphi}).

We also define a function
\be T(\theta, n) \eq \left[\cos \frac{N \theta}{2} \right]^{-n/N} \, \prod _{j=1}^n
\sin \left[  - \frac {\theta}{2} + \frac{\pi (2 j-1)}{2 N} \right] \period \ee

Consider a square lattice of $L$ sites. At each site $i$ there is a
``spin'' $\sigma_i$ which takes values $0,1, \ldots , N-1$. Adjacent 
horizontal sites $i, j$ (with $j$ to the right of $i$) interact with
Boltzmann weight function $ W_{pq} (\sigma_i - \sigma_j)$, and adjacent
vertical sites $i, j$ ($j$ above $i$) with
Boltzmann weight function $\overline{W}_{pq} (\sigma_i - \sigma_j)$.
The functions $W$, $\overline{W}$ are defined (to within normalization factors) by

\ba W_{pq} (n)/ W_{pq} (0)  &\eq &  T(\theta_q - \phi_p, n) /
 T(\theta_p - \phi_q, n) \comma \nonumber \\
\overline{W}_{pq} (n) / \overline{W}_{pq} (0)  &\eq &  T(\phi_p - \phi_q + \zeta,
n) / T(\theta_q - \theta_p - \zeta, n) \comma \ea
where \be \zeta = \pi /N \period \ee
The above relations ensure that the weight functions are periodic of period $N$:\\
$W_{pq}(n+N) = W_{pq}(n)$, $\overline{W}_{pq}(n+N) = \overline{W}_{pq}(n)$.

Define
\bd \rho_{pq} \eq \left\{ \prod_{n=0}^{N-1} W_{pq} (n) \right\} ^{1/N} \comma \ed
\be \Delta_{pq} \eq \left\{ {\rm det}_N \overline{W}_{pq} (i-j) \right\} ^{1/N}  \ee
i.e. $\Delta_{pq}^N$ is the determinant of the cyclic $N$ by $N$ matrix with
entry $\overline{W}_{pq} (i-j)$ in position $(i,j)$.

Let $Z$ be the partition function of the model. Then from (2.16), (3.22) and (3.42) of
\cite{freeenergy88}, the partition function per site is
\be \label{ZL}
Z^{1/L} \eq \rho_{pq} \; \Delta_{pq} \;  e^{-\Lambda_{pq}}  \ee
so $\Lambda_{pq}$ is the dimensionless free energy per site in the normalization in which
$\rho_{pq}$, $\Delta_{pq}$ are both one. This is the function $\Lambda_{pq}$ of 
\cite{freeenergy88}.

Now we consider the result (21) of \cite{RJB90}. There we used the normalization $\rho_{pq}
= \Delta_{pq} = 1$, and $V(t_q,\lambda_q)^{1/L}$ is the partition function per site, so the
lhs of (21) is $-4 N \Lambda_{pq}$.

We do need to reconcile the notations.
The variables $u_p, u_q$ of
\cite{RJB90} are the same as those above, except that they lie in the interval $-\pi  < u_p,
u_q < -\pi /2$.  We replace them by 
$u_p - \pi, u_q - \pi$ (and negate $v_p, v_q$). This is  equivalent to replacing the
rapidities
$p, q$ by $R^{-1}p, R^{-1} q$, which in turn is equivalent to merely rotating the lattice
through $180^{\circ}$, so does not change the free energy, but does ensure that the new
$u_p, u_q$ each lie in the interval (\ref{u-interval}), so we can directly compare the result
of
\cite{RJB90} with the equations of \cite{freeenergy88}.

Making these substitutions, eqn. (21) of
\cite{RJB90} gives, for $0 < u_p, u_q < \pi /2$, 
\ba \label{formula} 
\Lambda_{pq} \eq (\tau/2) \,  \log (\gamma_q/\gamma_p )  - 
  {\cal P} \! \int_{-\infty}^{\infty} \! \! \! \! \! & \!   &\! \! \!\! \! \!  e^{2 \beta
(u_q - u_p)/\pi }
\bigl\{ s(\beta) [G_p (\beta) G_q(-\beta) + {\rm cosech}^2 \beta] \, +  \nonumber \\
 & &  t(\beta) [G_p(\beta) + G_q(-\beta) ] \, \bigr\} d  \beta \comma \ea
where ${\cal P}$ denotes the principal-value integral,
\bd \tau = (N-1)/(2 N) \comma \ed
\be \label{defgamma}
J_p \eq [1+k^2 + 2 k \cos (u_p + v_p )]/{k'}^2 \eq \sin^2 (u_p + v_p ) /( {k'}^2 \sin^2
u_p )
\comma
\ee
\be \label{defs}  s(\beta ) \eq  [ N \, \sinh \beta \, \cosh (N-1)  \beta - \sinh N \beta ]/
[4N \beta \, \sinh N  \beta ] \comma \ee
\bd t(\beta ) \eq \sinh (N-1)  \beta /[4 \beta \, \sinh N  \beta ] \comma \ed
and
\be \label{Gpbeta}
G_p (\beta ) \eq \frac{\cos v_p}{\pi}
\int_{-\infty}^{\infty}\frac{e^{-\beta + 2
\beta (u_p + i x )/\pi} \, dx } 
{\sin (u_p + i x ) ( 1 + k^2 \sinh^2 x )^{1/2} } \period \ee
Note that negating $x$ is equivalent to complex conjugating the integrand in (\ref{Gpbeta}),
so $G_p (\beta)$ is a real function. Its value at $\beta = 0$ is
\be G_p(0) \eq 1 - 2 v_p /\pi \period \ee

In fact the integrals are convergent provided
$0 < u_p ,u_q < \pi $, so (\ref{formula}) should hold throughout this larger domain, each
side being analytic in $u_p, u_q$. For $u_p < u_q$ the Boltzmann weights are real and
positive and
$\Lambda_{pq}$ is the true free energy. For $u_p > u_q$ it is its analytic continuation. By 
negating $\beta$ it is easily seen that $\Lambda_{qp} = -\Lambda_{pq}$, in agreement with
(3.40) - (3.42) of \cite{freeenergy88}: this is the ``inversion relation''.

\section{Derivatives of $G_p(\beta)$}

The essential point of the method of \cite{freeenergy88} is that (because of the
star-triangle relation and the fact that $W_{pq}(n)$, $\overline{W}_{pq}(n)$ each depend
on only two parameters, rather than three) certain derivatives of $\Lambda_{pq}$ can be
expressed in terms of ``single-rapidity'' functions that depend on only $u_p$ or
$u_q$, but not both. (They also depend on $k$.) We now show that this follows directly from
(\ref{formula}).

To differentiate $G_p(\beta)$ with respect to $u_p$, keeping $k$ fixed, first change the
integration variable to $y$, where $x = y + i u_p$. Then differentiate, using (\ref{vu}),
and then change back to $x$ as the integration variable. The result is that the derivative
of $G_p(\beta)$ is equal to the rhs of (\ref{Gpbeta}), but with an extra factor in the
integrand:
\bd
- \, \frac{k^2 \sin u_p \cos u_p }{\cos^2 v_p} - \frac {i k^2 \sinh x \cosh x}
{1 + k^2 \sinh^2 x } \period \ed
(The first term comes from differentiating $\cos v_p$, the second from differentiating
the integrand.)

Expanding, this factor becomes
\be
 - \; \frac{k^2 \sin (u_p + i x ) \, [\cos u_p \cosh x + i {k'}^2 \sin u_p \sinh x]}
{\cos^2 v_p \, ( 1 + k^2 \sinh^2 x) } \period \ee

We see that $\sin (u_p + i x)$ appears in the numerator. This {\em cancels the same term in
the denominator in (\ref{Gpbeta}) }, leaving
\be \label{dGdu}
\frac{\partial}{\partial u_p }  G_p(\beta)  \eq
- \, \frac{k^2 e^{-\beta + 2 \beta u_p/\pi} }{ \cos v_p } \; [\cos u_p \, A(\beta) - {k'}^2
\sin u_p
\, B(\beta) ]
\comma \ee
where
\ba \label{AB}
A(\beta) & \eq & \frac{1}{\pi} \; \int_{-\infty}^{\infty} 
\frac{ \cos (2 \beta x/\pi ) \cosh x \; dx} 
{(1 + k^2 \sinh^2 x)^{3/2}}  \nonumber \\
B(\beta) & \eq & \frac{1}{\pi} \; \int_{-\infty}^{\infty} 
\frac{\sin (2 \beta x/ \pi ) \sinh x \; dx} 
{(1 + k^2 \sinh^2 x)^{3/2}}  \period \ea

We can also differentiate  $G_p(\beta)$ with respect to $k$, keeping $u_p$ fixed.
This time the extra factor in the integrand is
\bd
- \; \frac{k \sin^2 u_p }{\cos^2 v_p} - \frac { k \sinh^2 x }
{1 + k^2 \sinh^2 x } \eq - \; \frac { k \sin(u_p + i x) \sin ( u_p - i x) }
{ \cos^2 v_p \,  (1 + k^2 \sinh^2 x )} \period \ed
Again $\sin (u_p + i x)$ appears in the numerator, so cancels out of the full integrand,
leaving
\be\label{dGdk}
\frac{\partial}{\partial k}  G_p(\beta)  \eq - \; \frac{k \,  e^{-\beta + 2 \beta u_p/\pi
}}{\cos v_p}
\; [\sin u_p \, A(\beta) + \cos u_p \, B(\beta) ] \period \ee

 Note that $A(\beta)$, $B(\beta)$ are independent of $u_p$. Although the rapidity
parameter $u_p$ is locked into the integral expression (\ref{Gpbeta}) for $G_p(\beta)$,
it  can be taken outside the integrals occurring in the derivatives of $G_p(\beta)$. This
is the key to this calculation: indeed, most of what we do from now on is merely elementary
differentiation and straightforward (if cumbersome) algebraic manipulation.

\section{Derivatives of $\Lambda_{pq}$}

First differentiate $\Lambda_{pq}$ with respect to $k$, keeping
$u_p$ and $u_q$ fixed. Noting that
\bd
\frac{\partial}{\partial k} \; \log J_p 
\eq   2 \cos u_p /({k'}^2 \cos v_p)   \comma \ed
and using (\ref{dGdk}), we obtain
\be \label{derk}
 \frac{\partial}{\partial k} \Lambda_{pq} \eq 
\frac{ \ell_p  \sin u_q  + \emm_p \cos u_q -  \ell_q  \sin u_p -  \emm_q \cos u_p  }
{k \cos v_p \cos v_q}
 \comma \ee
where
\be \label{defellemm}
 \ell_p =  k^2 L_p \,\cos v_p  \sep \emm_p = 
\tau (k /{k'}^2) \cos v_p  - k^2 M_p \, \cos v_p  \comma \ee
$L_p, M_p$ being the integrals
\ba \label{LM}
L_p &  = & {\cal P} \int_{-\infty}^{\infty} e^{\beta -2 \beta u_p/\pi} \, 
[s(\beta) G_p(\beta) + t(\beta)] \, A(\beta) \; d \beta \comma \nonumber \\
M_p &  = &  \int_{-\infty}^{\infty} e^{\beta -2 \beta u_p/\pi}
\,  [s(\beta) G_p(\beta) + t(\beta)] \, B(\beta) \; d \beta \period \ea

Secondly, differentiating $\Lambda_{pq}$ with respect to $u_p$, $u_q$ and summing, noting
that
\bd 
\frac{\partial}{\partial u_p} \log J_p \eq -\frac{2 k \sin u_p }{ \cos v_p } \ed
 we obtain
\be \label{derupq}
\left( \frac{\partial}{\partial u_p} + \frac{\partial}{\partial u_q}
\right) \Lambda_{pq} \eq \frac{ \ell_p \cos u_q - {k'}^2  \emm_p \sin u_q -
 \ell_q  \cos u_p + {k'}^2 \emm_q \sin u_p }{ \cos v_p \cos v_q }
\period \ee
The contributions of the term $(\tau /2)\log(J_q/J_p)$ have been incorporated
into $\emm_p$.

Now define functions $x_p, y_p$ so that
\be \ell_p = (\epsilon  - x_p) \cos u_p + {k'}^2  y_p \sin u_p \sep
\emm_p =  (\epsilon -  x_p) \sin u_p -  y_p \cos u_p \period \ee
Here $\epsilon$ is a ``constant'' - independent of $u_p$ but dependent on $k$.
It is included with $x_p$ for later convenience.

Solving these equations for $x_p, y_p$ gives
\ba \label{xy}
 x_p & \eq &  \epsilon - (\ell_p \cos u_p + {k'}^2 \emm_p \sin u_p )/\cos^2 v_p
\comma
\nonumber \\  y_p & \eq & (\ell_p \sin u_p - \emm_p \cos u_p )/\cos^2 v_p \comma \ea
while substituting them into (\ref{derk}), (\ref{derupq}) gives
\be \label{derk1}
 \frac{\partial}{\partial k} \Lambda_{pq} \eq 
\frac{\sin (u_p + u_q) ( x_q - x_p) +c_{pq} (y_q - y_p)}
{k \cos v_p \cos v_q}
 \comma \ee

\be \label{derupq1}
\left( \frac{\partial}{\partial u_p} + \frac{\partial}{\partial u_q}
\right) \Lambda_{pq} \eq \frac{c_{pq} ( x_q - x_p) - 
{k'}^2 \sin (u_p + u_q) (y_q - y_p) }{\cos v_p \cos v_q }
\comma \ee
where \bd
c_{pq} \eq \cos u_p \cos u_q - {k'}^2 \sin u_p \sin u_q \period \ed

These are precisely the equations (3.45), (3.46) of \cite{freeenergy88}, so we have verified
that the double integral expression obtained in \cite{RJB90} has derivatives of this form,
and have (for the first time) obtained explicit expressions for the single-rapidity 
functions $x_p, y_p$ in \cite{freeenergy88}. 

Note that these functions $ x_p,  y_p$  are not to be confused with the
simple ratios
$a_p/d_p, b_p/c_p$ of the original rapidity parameters \cite{BPAY88}, elsewhere referred to
as $x_p, y_p$.

We shall find it helpful to regard $G_p(\beta),\ldots,  x_p, y_p$ as
functions $G(u_p,
\beta), \ldots,$ \\
$ x(u_p), y(u_p)$ of the variable
$u_p$, defining  $\cos v_p$ as the positive square root of $(1 - k^2 \sin^2 u_p )$ for $u_p$
real. The functions are defined by
the above integrals for $0 < u_p <
\pi$. Outside this interval and in the vicinity of the real axis they are defined by
analytic continuation.

From now on we drop the suffix $p$ and regard $u$ as an independent variable; $v$
is a dependent variable defined by (\ref{vu}), in particular
$\cos v = \surd ( 1 - k^2 \sin ^2 u )$. We refer to quantities that are independent of $u$
as ``constants'': they may still (and usually do) depend on $k$.

With these definitions, $G(u, \beta), \ldots, x(u), y(u)$ are real analytic
functions of $u$ for
$u$ real. In the complex plane they have branch points at 
\be \label{brpoints}
u = (n-1/2) \pi \pm \cosh^{-1} (1/k)  \ee
for all integers $n$.

Note that the function $G(u, \beta)$ of this paper is {\em not} the function $G(u)$ of
\cite{freeenergy88}. Nor should the function $x = x_p$ be confused with the integration
variable $x$ in (\ref{Gpbeta}) - (\ref{dGdk}).

\section{Differential equations for $x, y$}

An essential step in the working of \cite{freeenergy88} was the derivation of the  pair
of coupled partial differentiation equations (4.12) for $x(u)$ and $y(u)$:
\ba \label{4.12}
2 x + \left( \frac{\partial y}{\partial u} - k \frac{\partial x}{\partial k} \right)
\cos^2 v 
& = & \alpha - (\alpha + \delta) \sin^2 u \comma  \\
 -k^2 x \sin 2 u + \left( \frac{\partial x}{\partial u} + k {k'}^2
\frac{\partial y}{\partial k} - 2 y \right) \cos^2 v & = & \half ({k'}^2 \alpha + \delta)
\sin 2 u + \eta  \cos^2 v \period \nonumber
\ea
Here $\alpha,  \delta, \eta$ are unknown  constants. (This $\eta$ is the $\beta$ of
\cite{freeenergy88}.) 

Much of \cite{freeenergy88} is concerned with evaluating these
constants, but we  can now obtain explicit expressions for them. 
Using (\ref{defellemm}) - (\ref{xy}) to express $x, y$ in terms of $L, M$, and ignoring for
the moment the contribution of the constant term $\epsilon$ in the definition of $x(u)$, we
find that the left-hand sides of the two equations (\ref{4.12}) are, respectively,
\bd
 k^2  ( R \cos u + S \sin u )  \cos v  \sep
 k^2  ( {k'}^2 R \sin u - S \cos u ) \cos v \comma \ed
where
\ba \label{RS}  R & \eq &  L +  k \frac{\partial L}{\partial k} + \frac{\partial M}{\partial
u} 
\nonumber \\
S & \eq & \frac{\partial L}{\partial u} + (3 k^2 -1) M - k {k'}^2 \frac{\partial M}{\partial
k} \period \ea

From the definitions (\ref{AB}) we can establish
\ba \label{derAB}
 A(\beta) +  k \frac{\partial  A(\beta)}{\partial k} - 2 \beta B(\beta) /\pi & \eq & 0 \comma
\nonumber
\\  -2 \beta A(\beta)/\pi + (3 k^2 - 1) B(\beta) - k {k'}^2 \frac{\partial B(\beta)}
{\partial k}  & \eq & 0 
\period
\ea Substituting  the definitions (\ref{LM}) of $L(u)$ and $M(u)$ into (\ref{RS}) , and
noting that $s(\beta)$, $t(\beta)$ are independent of both $u$ and $k$, it follows that the
only contributions to $R$ and $S$ come from the derivatives of $G(u,\beta)$. Using
(\ref{dGdu}) and (\ref{dGdk}), we obtain
\be R = k^2 V \cos u \, /\cos v \sep S = k^2 {k'}^2  V \sin u \,  /\cos v \comma \ee
where
\be \label{defV}
V \eq 2 \int_{-\infty}^{\infty} s(\beta ) A(\beta) B(\beta) d \beta \period \ee
(There are also terms involving the integrals of $s(\beta) A^2 (\beta) $ and 
$s(\beta) B^2 (\beta) $, but these have odd integrands, so vanish.) Hence the left-hand
sides of (\ref{4.12}) are
\bd k^4  V[ -1 + (2-k^2) \sin^2 u]  \sep -k^4
{k'}^2 V \sin 2 u  \comma \ed respectively.
These are indeed of the same form as the rhs, and at this stage $\alpha, \delta, \eta$
have the values
$ - k^4 V$, $ - k^4 {k'}^2 V$, $0$, respectively.

We are still not quite in a position to directly compare our results with
\cite{freeenergy88}. There it was pointed out that that one can add arbitrary constants to
$x(u)$ and $y(u)$ without changing (\ref{derk1}), (\ref{derupq1}), and this freedom was used
to ensure that $\eta = 0$ and $\alpha + \delta = 0$. We already have $\eta = 0$, which
means we do not have to adjust $y(u)$. We do have to adjust
$x(u)$: we can do this by a suitable choice of $\epsilon$, which parameter gives
contributions $2 \epsilon - k \cos^2 v \; d \epsilon /dk$, $-k^2 \epsilon
\sin 2 u$ to the left-hand sides of (\ref{4.12}), and hence contributions
$2 \epsilon - k \; d \epsilon /dk$, $-2 \epsilon + k {k'}^2  \; d \epsilon
/dk$, $0$ to $\alpha$, $\delta$ and $\eta$. Altogether we therefore have
\be
\alpha + \delta \eq - k^4 V - k^4 {k'}^2 V - k^3 \; d \epsilon /dk \comma \ee
so to make $\alpha + \delta = 0$ we choose
\be
d \epsilon /d k \eq - k ( 1+{k'}^2) V \period \ee

If we define
\ba \label{defXY}
 X  & = & (\pi k^2 /2) \, \int_{-\infty}^{\infty} \beta^{-1} \, s(\beta)
A^2 (\beta) d
\beta \nonumber \\
 Y  &= & (\pi  k^2 {k'}^4  /2) \, \int_{-\infty}^{\infty} \beta^{-1} \, s(\beta) B^2 (\beta)
d
\beta \comma \ea
then from (\ref{derAB}) we can verify that
\be
\frac{d X}{d k} \eq k V \sep \frac{d Y}{ d k} \eq - \, k {k'}^2 V \comma \ee
so we can take
\be \label{epsXY}
\epsilon \eq Y - X \period \ee

This choice ensures that
\be
\lim_{u \rightarrow \pm i \infty} x(u) \eq 0 \comma \ee
and the final adjusted values of the constants in (\ref{4.12}) are
\be \label{eqalpha}
\alpha \eq - \, \delta \eq 2 [ k^2 {k'}^2 V + Y - X] \sep \eta \eq 0 \period \ee
The constant $\lambda$ that figures prominently in \cite{freeenergy88} is 
\be \label{deflambda}
\lambda \eq - \; \frac{k'}{k} \frac{d}{dk} \left( \frac{k' \alpha}{k^2} \right) 
\period \ee

\section{Properties of the functions}

\subsection*{Periodicity}

We should perform what checks we can to see if the functions $x(u), y(u)$ have the
properties used in \cite{freeenergy88}. Here we investigate their periodicity relations 
under the mapping $u \rightarrow u + \pi$. Here
$u_p$ and $u$ are the same variable.

First we have to analytically continue $G(u,\beta)$. As $u_p$ is increased beyond $\pi$ in
(\ref{Gpbeta}), the contour of integration has to shifted above the real axis so as to
remain above the pole at $x = i(u_p - \pi)$. The effect is the same as leaving the contour
on the real axis but adding the contribution from a small circle, traversed clockwise,
round this pole. This contribution is $e^{\beta}$. 

Thus for $\pi < u < 2 \pi$,   $G(u,\beta)$ is defined by (\ref{Gpbeta}), but with an
additional term $e^{\beta}$ on the rhs. It follows
that for $ 0 < u < \pi$, 
\be \label{perGpbeta}
G(u+\pi, \beta ) \eq   - e^{2 \beta } \, G(u,\beta) + 2 \; e^{\beta} \period \ee

Similarly writing $L_p, M_p$ as $L(u_p), M(u_p)$, and noting that
\be \label{st}
s(\beta) + t(\beta) \, \cosh \beta  \eq \tau/(2 \beta) \comma \ee
it follows from (\ref{LM}) that
\ba \label{LMper}
L(u) + L(u+\pi) & \eq & \tau \, {\cal P} \int_{-\infty}^{\infty}
e^{-2 \beta u /\pi } \, \beta^{-1} A(\beta) \, d\beta  \comma \nonumber \\
M(u) + M(u+\pi) & \eq & \tau \int_{-\infty}^{\infty}
e^{-2 \beta u /\pi } \, \beta^{-1} B(\beta) \, d\beta  \period \ea

The factors $\cos(2 \beta x/\pi), \; \sin(2 \beta x/\pi)$ in (\ref{AB}) can be written as the
real and imaginary parts of $\exp (2 i \beta x /\pi )$. Doing this, substituting the
resulting expressions for $A(\beta), B(\beta)$ into (\ref{LMper}),  differentiating with
respect to $u$ or integrating by parts in the expression for $B(\beta)$, we can arrange
that the $\beta$ integration gives a delta function. The $x$ integration can then be
performed immediately, giving
\ba
L(u) + L(u+\pi) & \eq & -  2 \tau \; \sin u / \cos v  \comma \nonumber \\
M(u) + M(u+\pi) & \eq &  \frac {2 \tau }{{k'}^2} \, \left[ \frac{1}{k} -
  \frac{\cos u}{\cos v} \right] \period \ea
It follows that
\ba \ell (u) + \ell (u + \pi ) & \eq & - 2 \tau  k^2 \sin u  \comma \nonumber \\
\emm (u) + \emm (u + \pi ) & \eq &   2 \tau ( k^2 /{k'}^2 ) \cos u    \ea

and 
\be
x(u+\pi) = x(u) \sep y(u+\pi) = y(u) +  2 \tau  k^2 / {k'}^2 \period \ee
 Thus 
$x, dy/du$ are strictly periodic functions of $u$, of period $\pi$.

We have derived these periodicity relations for $0 < u < \pi$, but by analytic continuation
they must be true for all real $u$.

\subsection*{Evenness/oddness}

We can replace $u_p$ in (\ref{Gpbeta}) by $\pi - u_p$ without leaving the domain of
validity of the equation. Doing so, then negating
$x$, we obtain
\be
G(\pi - u, \beta ) \eq G(u,- \beta) \comma \ee
from which it follows that $M(u), \emm(u), x(u)$ are unchanged by replacing
$u$ by $\pi - u$, while $L(u), \ell(u), y(u)$ are negated. Using the above periodicity
relations, we obtain
\be \label{evenodd}
x(-u) = x(u) \sep y(-u) + y(u) = - 2 \tau  k^2/{k'}^2 \period \ee
Thus $x(u)$ is an even function, $y(u) +  \tau k^2/{k'}^2$ is odd.

The above  periodicity and evenness/oddness properties agree with those given in eqns
(5.3), (5.4), (6.9) of \cite{freeenergy88}.

\section{The low-temperature limit $k \rightarrow 1$}

When $k \rightarrow 1$, the branch points (\ref{brpoints}) pinch onto the real axis
and the functions have discontinuities in their derivatives at $u = (n-1/2) \pi$.
If $-\pi/2 < u < \pi/2$, then  $\cos v = \cos u$ and (analytically continuing from $0 <
u < \pi/2$) we obtain 
\bd G(u,\beta ) \eq \left\{ 1 - e^{-\beta + 2 \beta u_p/\pi} \right\} /\sinh \beta \comma
\ed
\bd
A( \beta ) = 2 \beta /( \pi \sinh \beta ) \sep 
B( \beta ) =  2 \beta^2 /(\pi^2 \sinh \beta ) \period \ed
We also find that the constants $\epsilon, X, Y$ are
\bd
\epsilon \eq - X \eq - \, \frac{1}{2N^2}  \; \sum_{j=1}^{N-1} \, (N\! - \!2 j) \cot
(\pi j / N)  \sep Y \eq 0 \period \ed
Evaluating $L(u)$, $\ell(u)$, ${k'}^2 \emm(u)$, using the identity
\be \tan u \eq \frac{1}{N} \; \sum_{j=1}^N \cot \frac{(j\! - \! 1/2) \pi -
u}{N} \comma \ee
we find that
\be \label{xpk=1}
x(u) =  \frac{1}{2 N^2} \; \sum_{j=1}^{N} \, (2 j \! - \! N \! - \! 1) \cot
\frac{(j\! - \! 1/2) \pi - u}{N} \sep
 y(u)  = - \tau /{k'}^2 \period \ee

For $\pi/2 < u < 3\pi/2$ we  note that $\cos v$ is then $-\cos u$. Some of the terms
in the above working are then negated and we obtain 
\be
x(u) = \frac{1}{2 N^2} \; \sum_{j=2}^{N+1} \, (2 j \! - \! N \! - \! 3) \cot
\frac{(j\! - \! 1/2) \pi - u}{N} \sep  y(u) =  \tau /{k'}^2  \comma \ee

These results are consistent with the above
periodicity and evenness properties.
 Equations (3.25b) and  (5.2) of
\cite{freeenergy88} are in error in that the rhs of both should be divided by 2.  With this
correction, (5.2) and (4.13) therein are consistent with  (\ref{xpk=1}).

For $-\pi/2 < u < \pi/2$, we can also write (\ref{xpk=1}) as
\be x(u) \eq - ( 2N \cos u )^{-1}  \, \sum_{r=1}^{N-1} \frac{\cos (u - 2 r u /N)}
{ \sin (\pi r/N) } \comma \ee
from which we can verify  that eqn. (5.12) of \cite{freeenergy88} is
correct as written. Fortunately it is this equation that is subsequently used in
\cite{freeenergy88}, rather than (3.25b) or (5.2), so there is no reason to doubt
the results of that paper. 

In this limit we can verify that
\be \lambda \eq 2 \epsilon \comma \ee
in agreement with (5.25) of \cite{freeenergy88}.

\section{The critical case: $k \rightarrow 0$}

When $k \rightarrow 0$ the model becomes critical. To investigate the behaviour in this
limit it is helpful to note that $A(\beta)$ and $B(\beta)$ can be expressed in terms of 
hypergeometric functions.
Let
\be 
J(m,n) \eq \int_{-\infty}^{\infty} \frac{e^{2imx} \, dx }{(1 + k^2 \sinh^2 x)^{n}} \sep k =
1/\cosh \theta \period \ee
Then, taking $\theta$ to be real, we can establish that
\be \label{identJ}
J(m,n) \eq \half \, e^{2im\theta} \, (1 + e^{2 \theta})^{2n} \, B(n  + i m, n - i m) \; 
F(n+ i m, n; 2 n; 1-e^{4 \theta} ) \comma \ee
provided $n$ is real and positive and $| {\rm Im} (m) | < n$.
Here $B(m,n) = \Gamma (m) \Gamma (n) /\Gamma (m+n)$ is the beta function and
$F(\alpha,\beta;\gamma;z)$ is the usual hypergeometric function. We can use standard
transformation formulae (9.131 of
\cite{GR}) to verify that the rhs of (\ref{identJ}) is unchanged by negating either $m$ or
$\theta$.

Using formula (9.131.2) of \cite{GR}, it follows that
in the limit when
$k \rightarrow 0$ and $\theta \rightarrow +\infty$,
\be J(m,n) \rightarrow \half \left\{ e^{2 i m \theta} \Gamma (n - i m) \Gamma (im) +
e^{-2im\theta} \Gamma (n + i m) \Gamma (-i m) \right\}/\Gamma (n)  \ee
(neglecting terms of relative order $e^{-2 \theta}$ or smaller).

The functions $A(\beta)$, $B(\beta)$ defined in (\ref{AB}) can be written as sums and
differences of functions $J(m,3/2)$.
Define
\be c_+ (\beta) \eq  \pi^{-3/2} \, \Gamma(1+ i\beta/\pi) \Gamma (\half - i\beta/\pi)
\sep c_- (\beta) \eq  \pi^{-3/2} \, \Gamma(1- i \beta/\pi) \Gamma (\half +  i \beta/\pi) \period
\ee Then we find that
\ba \label{ABk0}
A(\beta) & \eq & k^{-1} \, \left[
e^{-2 i \beta \theta/\pi} \, c_+ (\beta) + 
e^{2 i \beta \theta/\pi} \, c_- (\beta)  \right] \comma
\nonumber \\ B(\beta) & \eq &  k^{-1} \,  \left[
i \, e^{-2 i \beta \theta/\pi} \,  c_+ (\beta) - i
\, e^{2 i \beta \theta/\pi} \,  c_- (\beta) 
\right] \period
\ea 

For $N = 2$, we see from (\ref{defs}) that $s(\beta ) = 0$, so $\epsilon$, $\alpha$,
$\lambda$ are zero. For $N > 2$, using (\ref{defXY}),  (\ref{epsXY}) we can write
$\epsilon$ as an integral over a quadratic form in  $A(\beta),
B(\beta)$. Substituting the forms (\ref{ABk0}),  the explicit $k$ factors cancel, we obtain
terms proportional to  $e^{-4i\beta \theta/\pi}$ and $e^{4i\beta \theta/\pi}$, and cross
terms where such exponentials have cancelled out. To leading order (for $k$ small) the
cross terms cancel, so their contribution to $\epsilon$ is at most of order $k^2$. The other
two terms differ only by negating $\beta$, so each give the same contribution to $\alpha$
and we obtain
\be \label{epsilonint}
\epsilon \eq - 2 \pi \, \int_{-\infty}^{\infty} s(\beta) \, e^{4i\beta
\theta/\pi}
\, \beta^{-1}  \; [ c_- (\beta) ]^2  \, d \beta \period \ee
We shall also need $\alpha$: this is of the same form as $\epsilon$, but with an extra
factor  $(2  + 4 i \beta/\pi)$ in the integrand.

 The  integrand in (\ref{epsilonint}) is is an analytic function
of $\beta$ on the real axis. The nearest singularity in the upper half-plane is 
a pole at $\beta = i \pi/N$, coming from $s(\beta)$, so in the
limit of $\theta$ large, we can close round the upper half plane and the results will be
dominated by the contribution from this pole. The residue of $s(\beta)$ is $(1/8 \pi) \sin 2
\pi /N$, so we obtain
\be \epsilon \eq - r k^{4/N}/(N-2) \sep \alpha = -2 r k^{4/N}/N \comma \ee
where (using standard formulae for the Gamma function)
\ba r & = & (2 \pi^3)^{-1} \, 2^{-4/N} \; N (N-2) \; \Gamma^2 (1 +{\textstyle\frac{1}{N}} ) 
\Gamma^2 ( \half - {\textstyle
\frac{1}{N}} ) \sin (2 \pi /N) \nonumber \\
& = & (N-2) \, \tan (\pi /N) \; \Gamma^4 (1/N) /[4 \pi^2 N \, \Gamma^2 (2/N)]
\period \ea
Hence from (\ref{deflambda}), to leading order (for $N > 2$),
\be \lambda \eq - 4 (N-2) \, r \, k^{4(1-N)/N}/N^2 \comma \ee
in agreement with
eqns. (6.2) -- (6.4) of \cite{freeenergy88}.

From (\ref{dGdu}) and (\ref{dGdk}), neglecting terms of relative order $k^2$,
\ba \label{expG}
G(u, \beta) \eq {\rm sech \;} \beta \; - \;  i k e^{-\beta+2 \beta u /\pi} 
\Bigr\{
e^{-iu - 2 i \beta \theta/\pi} \,(1+2 i \beta/\pi)^{-1} \,  c_+ (\beta) \;  - &&
\nonumber
\\
 e^{iu + 2 i \beta \theta/\pi} \, (1-2 i \beta/\pi)^{-1} \, c_- (\beta) 
\Bigr\} && \period \ea
Now consider the functions $L(u), M(u)$ defined by (\ref{LM}) as integrals over $\beta$.
Substituting the above forms of $A(\beta), B(\beta), G(u,\beta)$, we obtain three distinct
contributions.

\paragraph{(i)  Terms coming from the ${\rm sech \,}\beta$ component of $G(u, \beta)$:} 

Using (\ref{st}), these correspond to replacing $s(\beta) G_p(\beta) + t(\beta)$ in
(\ref{LM}) by $\tau/(2 \beta \cosh \beta )$. Their only dependence on $N$ is via the 
external factor $\tau$, i.e. they are proportional to {{$(N-1)/N$}}. Using (\ref{ABk0}),
taking the $\beta$-integral to be principal-valued,  and negating $\beta$ in the terms involving
$c_+(\beta)$, the integrand contains a factor
$e^{2 i \beta \theta/\pi}$. We can close round the upper half-plane and in the limit
$\theta \rightarrow +\infty $ the dominant contributions are from a possible simple pole at
$\beta = 0$ and a double pole at $\beta = \pi/2$. The first gives contributions $0, \tau/k$
to $L(u), M(u)$, respectively  - these cancel the $\tau (k/{k'}^2) \cos v$ term in
(\ref{defellemm}). The second  gives contributions
$(d_1+2
\tau
\theta/\pi)
\cos u -
\tau (1-2u/\pi) \sin u$, $-(d_1+2 \tau \theta/\pi) \sin u - \tau
(1-2u/\pi) \cos u$, where $d_1$ is independent of both $u$ and $k$.

\paragraph{(ii) Cross terms:}

The other contributions come from the terms in (\ref{expG}) involving $c_+(\beta),
c_-(\beta)$. They contain $s(\beta)$ as a factor in the integrand, so only occur for $N > 2$. 
Using (\ref{ABk0}), they can naturally be grouped into three terms: those proportional to
$e^{-4i\beta
\theta/\pi} c_+(\beta)^2$,
$c_+(\beta) c_-(\beta)$, and
 $e^{4i\beta \theta/\pi} c_-(\beta)^2$. The middle terms are cross terms, independent of
$\theta$ and $k$, and give contributions $d_2 \cos u$, $- d_2 \sin u$, to $L(u), M(u)$,
respectively, where $d_2$, like $d_1$, is independent of both $u$ and $k$.

\paragraph{(iii) Terms proportional to fractional powers of $k$:}

Finally there are the first and third terms mentioned in (ii). We negate the variable $\beta$
in the first term, which puts it into a form similar to the third.
The resulting integral can be closed round the upper half $\beta$ plane. There is no pole at the
origin and the nearest pole to the real axis is a simple pole of $s(\beta)$ at $\beta = i \pi
/N$. The calculation is now very similar to that of $\epsilon$: evaluating the residue we find
that the combined  contribution of these terms to $L(u), M(u)$ is $- r k^{4/N} (N^2 - 4)^{-1}
\cos u
$,  $- r k^{4/N} (N^2 - 4)^{-1}  \sin u $, respectively.

Substituting these various contributions into (\ref{defellemm}) and (\ref{xy}) and working to
order $k^2 \log k$ in terms in $x(u)$ that are independent of $u$, to order $k^{2+4/N}$ in
other terms, we obtain
\ba x(u)  & = & 2 \pi^{-1} \tau k^2 \log (k/2) - \frac{r k^{4/N}}{N-2} + \frac{r k^{2+4/N}
\cos 2 u}{N^2 - 4} \nonumber \\
&& \nonumber \\
y(u) & = & - \, \tau (1-2 u /\pi) k^2 - \frac{r k^{2+4/N} \sin 2 u}{N^2 - 4} \period \ea
These results agree with (and slightly extend) eqns. (6.6), (6.8) of \cite{freeenergy88}.

\subsection*{The scaling region}

In an earlier paper \cite{RJB96} we obtained the free energy and the functions $\lambda, x(u),
y(u)$ in the scaling region near $k = 0$, considerably extending the above results in a series
expansion in powers of $k^{4/N}$. We do not with to fully re-derive these results here, but
note that if we include not just the contributions discussed above from poles at $\beta =
i\pi/N$, but more generally at $\beta = i  j \pi/N$, where $j = 1,2, \ldots$, then
\bd \epsilon = - \sum k^{4j/N} \, r_j /(N-2j) \sep \alpha = - (2/N) \sum k^{4 j/N} \, r_j \ed
\be
\lambda = - (4/N^2) \, \sum (N-2 j) k^{4j/N - 4} \, r_j \comma \ee
where
\ba
r_j & = & \frac{N(N-2j)}{2 \pi^3 j^2 \, 2^{4j/N} } \, \sin \left( {\textstyle \frac{2 \pi j}{N}}
\right)
\Gamma^2 \left(1+{\textstyle \frac{j}{N}} \right) 
\Gamma^2 \left(\half -{\textstyle \frac{j}{N}} \right)  \nonumber \\
& & \nonumber \\
& = &
\frac{N-2j}{4 \pi^2 N} \; \tan ( \pi j/N )  \; \Gamma^4 (j/N) /\Gamma^2 (2 j/N)
\period \ea
The sums are over positive integer values of $j$ and we should restrict $j$ to be less than
$N/2$, since for greater values the contributions are smaller than other terms, of relative order
$k^2$, that we are ignoring.

Similarly, considering the contributions of such a pole to $L(u), M(u), x(u), y(u)$, the term
$r k^{2 + 4/N}/(N^2-4)$ above is replaced by $j \, r_j \, k^{2 + 4j/N}/(N^2-4j^2)$ and we obtain
\ba x(u)  & = & 2 \pi^{-1} \tau k^2 \log (k/2) - \frac{r k^{4/N}}{N-2} + \sum \frac{j \, r_j
\, k^{2+4j/N}
\cos 2 u}{N^2 - 4j^2} \nonumber \\
&& \nonumber \\
y(u) & = & - \, \tau (1-2 u /\pi) k^2 - \sum \frac{j \, r_j \, k^{2+4j/N} \sin 2 u}{N^2 - 4 j^2}
\period
\ea
These results agree with eqn. (B2) of \cite{RJB96}, and with the relevant $H_1 (k)$ term in eqns.
(B5), (B6).

\section{The case $N = 2$}

When $N = 2$ the model reduces to the Ising model. As remarked above, the function $s(\beta)$
then vanishes, causing some simplifications, in particular $\epsilon = $ $\alpha = $ $ \lambda =
$ $ 0$. Substituting (\ref{AB}) into (\ref{LM}), performing the $\beta$ integration, then
integrating by parts in the $x$-variable, we obtain, for $0 < u < \pi$,
\ba
L(u) & = & \frac{i}{4 \pi} \, \int_{-\infty}^{\infty} \frac{\sinh x \; dx }{\sin (u+i x) (1+k^2
\sinh^2 x )^{1/2} }\comma \nonumber \\
&&\nonumber \\
M(u) & = & \frac{1}{4 k {k'}^2} - \frac{1}{4 \pi {k'}^2} \, \int_{-\infty}^{\infty} 
\frac{\cosh x \; dx }{\sin (u+i x) (1+k^2
\sinh^2 x )^{1/2} }\period \ea

Substituting these into (\ref{defellemm}), (\ref{xy}), the factor $\sin (u + i x)$ cancels out of
the expression for $x(u)$, leaving, in agreement with  (5.38) of \cite{freeenergy88},
\be
x(u) \eq - \, k^2 K'/\left( 2 \pi \cos v \right) \comma \ee
where
\be
K' \eq \int_{0}^{\infty} (1+ k^2 \sinh^2 x)^{-1/2} \; dx \ee
is the usual complete elliptic integral. 

For $y(u)$ we obtain
\be \label{y1}
y(u) \eq - \; \frac{ k^2}{4 \pi {k'}^2 \cos v} \, \int_{-\infty}^{\infty} \frac{\cos u \cosh x -
i {k'}^2 \sin u \sinh x  }{\sin (u+i x) \surd (1 + k^2 \sinh^2 x ) } \, dx \period \ee
To handle this we introduce Jacobi's elliptic functions of modulus $k$. Define $w, t$ by
\be
\sin u = \sn w \sep \cos u = \cn w \sep \sinh x = -i \sn t \sep \cosh x = \cn t \comma \ee
where $0 < w < 2 K$ and $t$ is pure imaginary. Then (\ref{y1}) becomes
\be \label{y2}
y(u) \eq  \frac{i k^2}{8 \pi {k'}^2} \, \int_{- i K'}^{i K'} {\cal J}(w,t) \, dt \comma \ee
where
\bd {\cal J}(w,t) \eq \frac{2 \, ( \!\cn w \cn t - {k'}^2 \sn w \sn t ) }
{\! \dn w \, ( \sn w \cn t + \cn w \sn t  ) } \period \ed
By considering the poles and residues of ${\cal J}(w,t)$ as a function of $t$, 
we can verify that
\be {\cal J}(w,t) \eq 
 \left( \frac{H'}{H} + \frac{H_1'}{H_1} \right) \! \left( \frac{w + t}{2} \right) \, +  \,
 \left( \frac{\Theta'}{\Theta} + \frac{\Theta_1'}{\Theta_1} \right) \! \left( \frac{w - t}{2}
\right) \, - \, 2 \; \frac{\Theta_1'}{\Theta_1} \left( w \right)
\comma \ee
where for instance the first bracketted term on the rhs is the sum of the logarithmic derivatives
of the Jacobi theta functions $H$ and $H_1$, evaluated with arguments $(w+t)/2$. We can now
integrate ${\cal J}(w,t)$ directly with respect to $t$ (taking care to obtain the correct
branches of complex logarithms), giving
\be \label{y3} y(u) \eq \frac{k^2}{2 \pi {k'}^2} \left\{ \frac{\pi (w-K) }{2 K} +
K' \, \frac{\Theta_1'(w)}{\Theta_1 (w)} \right\}  \period \ee
It seems that the expression in (5.38) of \cite{freeenergy88} for $y(u)$ is in error: it should
be multiplied by $2 k$ and the first occurrence of $w_p$ replaced by $w_p - K$. When that is
done, it agrees with (\ref{y3}).

We have checked that these $N = 2$ results have the periodicity and eveness/oddness properties
discussed above, and are consistent with the $k = 1$ and $k \rightarrow 0$ limiting cases.

\section{The inversion and rotation relations}

We have already remarked at the end of section 1 that
\be \label{invLambda}
\Lambda_{qp} \eq - \Lambda_{pq} \comma \ee
and that this is the ``inversion'' relation \cite{RJB82} of the chiral Potts model.
Let $R$ be the automorphism such that $u_{Rp} = u_p + \pi$,
 $\; v_{Rp} = - v_p$. Then, using  ref. \cite{BPAY88},
\be W_{q,Rp}(n) \eq {\overline{W}}_{pq}(n) \sep \overline{W}_{q,Rp}(n) \eq
W_{pq}(-n) \comma \ee
so replacing $p, q$ by $q, Rp$ is equivalent to merely rotating the lattice through
$90^{\circ}$.  This leaves the partition function
$Z$ is unchanged. From (\ref{ZL}) it follows that
\be \label{rotnLambda}
\Lambda_{q,Rp} - \Lambda_{pq} \eq \log f_{q,Rp} - \log f_{pq} \comma \ee
where \be
f_{pq} \eq \left\{ {\rm det}_N \overline{W}_{pq} (i-j) \right\} ^{1/N} {\bigg/} 
\left\{ \prod_{n=0}^{N-1} \overline{W}_{pq} (n) \right\} ^{1/N} \period \ee
The aim of this section is to explicitly show that the formula (\ref{formula} ) satisfies
(\ref{rotnLambda}).

We first need to write $f_{pq}$ (or rather its logarithm) in an appropriate form. It is given
explicitly in eqn. (3.22) of
\cite{freeenergy88}. Using (3.33) therein, we can write it as 
\be \label{formfpq}
f_{pq} \eq N^{1/2} \, (h_{pq} /k')^{(N-1)/2N} \; g_{pq} \comma \ee
where
\ba \label{defgpq}
g_{pq} & = & \prod_{j=0}^{N-1} \{ 2 \sin[(u_q - u_p + \pi j)/N] \}^{(2
j+1-N)/2N} 
 \comma  \\
h_{pq} & = & 
\left\{ \frac{\cos [N(\theta_p - \phi_q)/2] \; \cos [N(\theta_q - \phi_p)/2]}
{\sin [N(\theta_q -
\theta_p)/2] \; \sin [N(\phi_q - \phi_p)/2]}
 \right\}^{1/2} \period \nonumber \ea

If $u_p, u_q$ are real and $0 < u_q - u_p < \pi$, then all the Boltzmann weights $W_{pq}(n)$,
 $\overline{W}_{pq}(n)$ are real and positive, and so are $g_{pq}, h_{pq}$. Using the relations
(\ref{thetaphi}), (\ref{vu}), we can write $h_{pq}$ as
\be
h_{pq} \eq \frac
{\sin ( N \theta_p ) \cos [N(\theta_q - \phi_p)/2] }
{{k'} \sin u_p \; \sin[N(\theta_q - \theta_p)/2]} \period \ee
Let us write the variables $x_p, y_p, \lambda_p$ of \cite{RJB90, Seoul} as
$\tilde{x}_p, \tilde{y}_p, \lambda_p$. Then, using the definition (\ref{defgamma})
of $J_p$,
\bd
\tilde{x}_p = e^{i \phi_p} \sep \tilde{y}_p = e^{i(\theta_p + \pi/N)} \comma \ed
\bd k \, {\tilde{x}}_p^N = 1 - k' /\lambda_p \sep  k \, {\tilde{y}}_p^N = 1 - k' \lambda_p
\sep J_p =  \lambda_p^2 \, e^{-2 i v_p}  \comma \ed
\bd  \lambda_p + \lambda_p^{-1} = (1 + {k'}^2 + k^2 e^{2 i u_p} )/k'  \sep  \lambda_p -
\lambda_p^{-1} = 2 ( k/k' ) \, e^{i u_p} \, \cos v_p \ed
and
\be \label{formhpq}
h_{pq} \eq i (1 - \lambda_p \lambda_q) /(\lambda_p - \lambda_q) \period \ee

Using the identity (56) of \cite{Seoul}, we can establish that
\be
\log g_{pq} \eq - {\cal P} \; \int_{-\infty}^{\infty} e^{-\beta + 2 \beta (u_q - u_p)/\pi}
[t(\beta) + s(\beta) \cosh \beta /\sinh^2 \beta ]  \, d \beta \ee
provided $0 < {\rm Re \,} (u_q - u_p ) < \pi$.

Consider $h_{pq}$ as a function of $u_p$, keeping $k$ and $u_q$ fixed. Differentiating and
using the above relations, we obtain
\be \label{dloghpq}
\frac{d \, \log h_{pq}}{d  u_p} \eq \frac{ \cos v_q }
{\sin (u_q - u_p) \cos v_p } \period \ee

Provided $0 < u_q < \pi$, it follows that $d \, \log  h_{pq} /d  u_p$ is analytic in a
vertical strip in the complex $u_p$ plane containing the imaginary axis, and tends exponentially
to zero as $u_p \rightarrow \pm i \infty$. It can therefore be Fourier analyzed in this strip.
Doing this and using the definition (\ref{Gpbeta}), we find that
\be \label{derivhpq}
\frac{d \, \log h_{pq}}{d  u_p} \eq \frac{1}{\pi} \; \int_{-\infty}^{\infty} 
\, e^{-\beta + 2\beta(u_q - u_p)/\pi} \, G_q (-\beta) \, d \beta  \ee
provided $-\pi/2 < {\rm Re\,} u_p < \pi/2$ and $0 < {\rm Re \,} (u_q - u_p ) < \pi$.

When $u_p \rightarrow + i \infty$, then $v_p \rightarrow + i \infty$, $\lambda_p \rightarrow
1/k'$ and $h_{pq} \rightarrow  i \lambda_q e^{-2i v_q}$.
When $u_p \rightarrow - i \infty$, then $v_p \rightarrow - i \infty$, $\lambda_p \rightarrow
\infty $ and $h_{pq} \rightarrow  - i \lambda_q$. Integrating (\ref{derivhpq}) with respect to 
$u_p$, it follows that
\be \label{loghpq}
\log h_{pq} \eq \frac{1}{2} \log J_q -
 {\cal P} \, \int_{-\infty}^{\infty} \frac{
e^{-\beta + 2\beta(u_q - u_p)/\pi} }{ 2 \beta } \, G_q (-\beta) \, d \beta \period \ee

We have introduced several complex variables, but if $u_p, u_q$ are real and $0 < u_q - u_p <
\pi$, then we recall that
$G_q(-\beta)$ is a real function, and
$h_{pq}$, $J_q$ are real and positive: (\ref{dloghpq}),
(\ref{derivhpq}), (\ref{loghpq}) are then real equations. The additional conditions $-\pi/2 < u_p
< \pi/2$, $0 < u_q < \pi$ are needed to ensure convergence of the integrals as written.

Noting that $\lambda_{Rp} = \lambda_p^{-1}$, from (\ref{defgpq}) and (\ref{formhpq}) 
 we can verify that
\bd
g_{q,Rp} = 1/g_{pq} \sep h_{q,Rp} = 1/h_{pq} \comma \ed
so from (\ref{formfpq}) - (\ref{loghpq}),
\ba \label{difff}
\log f_{q,Rp}  -  \log f_{pq}  & \! = \! & -2 \, \log g_{pq} - 2 \tau \log h_{pq}  \nonumber \\
& = & - \tau \log  J_q    +  
{\cal P} \! \int_{-\infty}^{\infty}
e^{-\beta + 2\beta (u_q - u_p) /\pi } {\cal X}_q(\beta) \, d\beta \comma \ea
where
\be
{\cal X}_q(\beta) \eq \tau \, \beta^{-1} \,  G_q (-\beta)  +  2 \, t(\beta) +  2 \, s(\beta)
\cosh
\beta /\sinh^2 \beta  \period
\ee
Using (\ref{st}), we can write this last relation as
\be
{\cal X}_q(\beta) \eq 2 \, s(\beta) [G_q(-\beta) + \cosh \beta/\sinh^2 \beta ] + 2 \,
t(\beta) [1 +  G_q (-\beta) \cosh \beta ] \period \ee

We can now explicitly verify that the result (\ref{formula} ) of \cite{RJB90, Seoul} satisfies
the rotation relation (\ref{rotnLambda}). The relation (\ref{perGpbeta}) 
implies
\be 
G_{Rp}(\beta) \eq - e^{2\beta} \, G_p(\beta) + 2\, e^{\beta} \period \ee
Using this identity and  (\ref{formula} ), and noting that $J_{Rp} = 1 / J_p$,  we find
\be
- \Lambda_{Rp,q} - \Lambda_{pq}  =  - \tau \log J_q  + 
{\cal P} \! \int_{-\infty}^{\infty} e^{-\beta+2 \beta(u_q - u_p)/\pi} \, {\cal X}_q(\beta)
\comma \ee the function $G_p(\beta)$ having cancelled out of the integrand.
The RHS of this result is the same as that of (\ref{difff}), while from (\ref{invLambda}) the
LHS is $\Lambda_{q,Rp} - \Lambda_{pq}$: (\ref{rotnLambda}) is therefore satisfied.

\section{Summary}

We have obtained the explicit expressions (\ref{defellemm}) - (\ref{xy}) for $x_p, y_p$.
We have shown that they satisfy the differential relations (\ref{4.12}), and that they agree
with previous results (corrected where necessary) for the low-temperature case $k \rightarrow
1$, for the critical case $k \rightarrow 0$, and for the Ising case $N = 2$.

In section 9 we have also explicitly shown that the free energy satisfies the inversion
and rotation relations. For other models that enjoy the ``rapidity difference'' property, there
is a uniformizing substitution that makes it easy to derive the free energy from these 
relations and a simple analyticity assumption. We do not know how to do this for 
the chiral Potts model, nor do we know how to solve similar relations for the generalized
spontaneous magnetization \cite{RJB98a,RJB98b,RJB98c}. The presentation in section 9
may cast some light on this outstanding problem.

\end{document}